\documentclass[useAMS,usenatbib]{mn2e}

\title[Neutron-star magnetic field evolution]{Fermion zero-mode influence on 
neutron-star magnetic field evolution}
\author[P. B. Jones]{P. B. Jones\thanks{E-mail:
p.jones1@physics.ox.ac.uk}\\
Department of Physics, University of Oxford, Denys Wilkinson Building, \\
Keble Road, Oxford, OX1 3RH.}

\begin{document}

\date{}

\pagerange{} \pubyear{}

\maketitle

\label{firstpage}

\begin{abstract}
The quantum phenomena of spectral flow which has been observed in laboratory 
superfluids, such as $^{3}$He-B, controls the drift velocity of proton type II 
superconductor vortices in the liquid core of a neutron star and so determines 
the rate at which magnetic flux can be expelled from the core to the crust.  In 
the earliest and most active phases of the anomalous X-ray pulsars and soft-gamma 
repeaters, the rates are low and consistent with a large fraction of the active 
crustal flux not linking the core.  If normal neutrons are present in an 
appreciable core matter-density interval, the spectral flow force limits flux 
expulsion in cases of rapid spin-down, such as in the Crab pulsar or in the 
propeller phase of binary systems.
\end{abstract}

\begin{keywords}
stars: neutron - stars: magnetic fields - pulsars: general
\end{keywords}

\section{Introduction}

It is now well-established that evolution and decay of the internal magnetic 
field, rather than rotational energy, is the dominant source of the persistent 
and burst X-ray luminosities in certain hot young isolated neutron stars.  These 
are observed as the anomalous X-ray pulsars and soft-gamma repeaters, referred to 
as magnetars (for reviews see Harding \& Lai 2006; Woods \& Thompson 2006; Pons, 
Link, Miralles \& Geppert 2007).  In principle, a large part of the internal 
magnetic flux could be contained in the liquid core of the star at densities less 
than $\sim 2\rho_{0}$, where $\rho_{0} = 2.8 \times 10^{14}$ g cm$^{-3}$ is the 
nuclear density.  In this interval, many calculations of pairing and of the 
equation of state (Baldo, Burgio \& Schulze 1998; Heiselberg \& Hjorth-Jensen 
2000) show that the core consists of neutrons, protons, relativistic electrons 
and possibly muons, and that the neutrons and protons are superfluids, pairing in 
the $^{3}{\rm P}_{2} - ^{3}{\rm F}_{2}$ and $^{1}{\rm S}_{0}$ states, 
respectively.  The proton energy gaps found in these calculations indicate
a type II superconductor (Baym, Pethick \& Pines 1969) However, Baldo \& Schulze 
(2007) have since shown that the inclusion of more refinements, including 
three-body forces, reduces the energy gap to values consistent with type I 
superconductivity for certain proton densities.  It is also the case that other 
authors have more recently considered the possibility of type I superconductivity 
with various motivations (Sedrakian, Sedrakian \& Zharkov 1997; Buckley, 
Metlitski \& Zhitnitsky 2004; Sedrakian 2005; Alford, Good \& Reddy 2005; 
Charbonneau \& Zhitnitsky 2007; Alford \& Good 2008).  But it remains a 
reasonable assumption that type II superconductivity exists for at least part of 
the density interval we are considering here.

In the type II system, the magnetic flux is quantized in units of $\phi_{0} = 
hc/2e = 2.07 \times 10^{-7}$ G cm$^{2}$ and is confined to the cores of proton 
vortices (see, for example, Tinkham 1996).
Steady-state motion of the proton vortices is determined by balance of the Magnus 
force, electromagnetic interactions, external forces including buoyancy forces 
(Muslimov \& Tsygan 1985; Jones 2006) and interaction with vortices of the 
rotating neutron superfluid (Muslimov \& Tsygan 1985; Sauls 1989; Jones 1991; 
Ruderman, Zhu \& Chen 1998), and forces derived from interactions between the 
quasiparticles of the system.  The distinct populations of superfluid 
quasiparticles concerned are those localized in the cores of proton vortices and 
those delocalized in the continuum.  Previous vortex velocity calculations (Jones 
2006) assumed both the internal temperatures $T < 10^{8}$ K of typical radio 
pulsars and negligibly small neutron and proton quasiparticle number densities.  
Under these conditions, proton vortices and therefore the internal flux 
distribution could be quickly expelled to the core-crust boundary.

The low-temperature assumption is not appropriate for magnetars whose X-ray 
luminosities in the active phase, $\sim 10^{3} - 10^{4}$ yr after formation, are 
thought to be a consequence of plastic flow of the crust (with its frozen-in 
field) under very high magnetic stresses (Harding \& Lai 2006; Woods \& Thompson 
2006).  This dissipative process must maintain a large input of thermal energy to 
the inner crust and core of the star so that internal temperatures are likely to 
be much higher than those inferred for the isolated radio pulsars. 
Neglect of continuum superfluid quasiparticles is not justified at these higher 
temperatures unless it can be established that the neutron energy gap is large at 
all relevant wavenumbers.  Although early calculations gave such values (Amundsen 
\& {\O}stgaard 1985),
much work (see the review by Heiselberg \& Hjorth-Jensen 2000) finds a small 
angle-averaged $^{3}{\rm P}_{2} - ^{3}{\rm F}_{2}$ neutron energy gap, 
$\Delta_{n} \sim 0.1$ MeV, at the wavenumbers concerned.  More recently, Khodel, 
Khodel \& Clark (2001) also obtained similar small values for certain 
neutron-neutron interactions.  But there is also the possibility of pion 
condensation at a matter density $\sim 2\rho_{0}$.  Its effect has been 
investigated by Khodel et al (2004) and is to increase the gap in the neutron 
single-particle spectrum.  Nevertheless, we regard the existence of a small 
neutron superfluid gap, or possibly even a normal Fermi liquid, as being not 
improbable for some wavenumber bands within the density interval considered in 
this paper.   

A new development in condensed-matter physics, the phenomenon of spectral flow 
described by Volovik (1996; see also the review of Kopnin 2002), has greatly 
clarified and extended our understanding of the forces acting on moving vortices 
and has been confirmed by measurements on the $^{3}{\rm He}$-B phase (Bevan et al 
1997).  The present paper applies these ideas to show that forces generated by 
spectral flow limit the velocities of proton vortices in the outer core of hot 
young neutron stars to an extent that is consistent with a large fraction of the 
active flux not linking the core.

\section[]{Spectral flow}

The fermion zero-mode excitations concerned are those quasiparticles localized 
within an isolated vortex core at energies below $\Delta_{p}$ (referred to the 
chemical potential as zero).  Approximate eigenvalues (Caroli \& Matricon 1965; 
de Gennes 1966) in the limit $p_{F}\xi \gg 1$, where $\xi$ is the superfluid 
coherence length and $p_{F}$ the Fermi wavenumber, are $E_{p\mu}$, where $\mu$ is 
the (half-integer) angular momentum component parallel with the vortex axis 
$\hat{\bf z}$, and
\begin{eqnarray}
E_{p\mu}  & = &  -\mu\hbar\omega_{0,}    \nonumber  \\
\hbar\omega_{0}& =  &  \frac{\hbar^{2}p_{F}}{\pi m^{*}_{p}\xi^{2}p_{\perp}} = 
	\frac{\pi\Delta^{2}_{p}p_{F}}{2E_{Fp}p_{\perp}} \ll \Delta_{p}.
\end{eqnarray}
(This expression neglects interaction with the core magnetic flux; see Bardeen et 
al 1969).
The proton bare and effective masses are $m_{p}$ and $m^{*}_{p}$ and its
Fermi energy is $E_{Fp}$. The components of the proton wave-vector parallel with 
and perpendicular to the vortex axis are $p$ and $p_{\perp}$. Values of $\mu$ are 
large at the temperatures of Table 1, so that localized quasiparticles 
interacting with an external heat-bath (consisting, in this case, of a 
distribution of continuum quasiparticles) can be treated by a Boltzmann equation 
(see Kopnin 2002) in which
the quantum number $\mu$ is replaced by a continuous variable.  In this 
semi-classical model, the vortex structure is that of a cylindrical hole in the 
superfluid order parameter and quasiparticle states within this hole are normal 
particle and hole orbits in the form of chords limited at each end by Andreev 
reflection (Andreev 1964), the specific process of reflection that occurs at a 
normal-superfluid interface.

The perturbation of the Boltzmann distribution function caused by vortex motion
consists of a displacement of the Fermi sphere in the plane perpendicular to 
$\hat{\bf z}$ which also contains the vortex velocity ${\bf v}_{L}$.
It relaxes through binary collisions with either localized (vortex-core) 
quasiparticles or with continuum quasiparticles.  Thus the relaxation rate is
given by the transport relaxation times,
 $\tau^{-1} = \tau^{-1}_{loc} + \tau^{-1}_{c}$, but only the latter process 
transfers momentum between the vortex and the continuum.  The existence of these 
two distinct relaxation processes means that, within the semi-classical model, 
the localized quasiparticles mediating the force between the continuum heat-bath 
with bulk velocity ${\bf v}_{T}$ and a vortex of velocity ${\bf v}_{L}$  have a 
steady-state bulk velocity ${\bf v}_{loc} = {\bf v}_{L}\tau/\tau_{loc} + {\bf 
v}_{T}\tau/\tau_{c}$.  This lags behind ${\bf v}_{L}$ in general, and in the 
limits $\tau_{c}$ or $\tau_{loc} \rightarrow \infty$ tends, as expected, to ${\bf 
v}_{L}$ or ${\bf v}_{T}$.  Thus spectral flow (Volovik 1996) depends on the 
velocity difference
\begin{eqnarray}
\tilde{\bf v} = {\bf v}_{L} - {\bf v}_{loc} = ({\bf v}_{L} - {\bf 
v}_{T})\tau/\tau_{c}.
\end{eqnarray}
In Andreev reflection, a particle (hole) is retroreflected as a hole (particle), 
so maintaining the orientation of the chord. It has angular momentum $\mu = 
[({\bf r} - \tilde{\bf v}t)\times {\bf p}_{\perp}]\cdot\hat{\bf z}$, where ${\bf 
r} - \tilde{\bf v}t$ is its coordinate in the vortex rest frame.  The 
retroreflection is not exact owing to the motion of the superfluid so that the 
chord precesses about the vortex axis with angular frequency $\omega_{0}$ in a 
direction opposite to that of the circulating superfluid (Stone 1996).  Motion of 
the vortex induces a finite rate of change of orbit angular momentum, given by 
the classical variable $\dot{\mu} = -[\tilde{\bf v}\times {\bf 
p}_{\perp}]\cdot\hat{\bf z}$, that drives spectral flow, which is the movement of 
quasiparticles both to and from the negative energy continuum $E_{p\mu} < 0$ 
along the continuous dispersion relation given by equation (1).  For $-\mu \sim 
10$ or more, we rely on the correspondence principle in assuming that 
classical-orbit quasiparticles give an adequate description of spectral flow.  In 
the case of the physical discrete spectrum, with half-integer $\mu$, the process 
of spectral flow relies on the existence of sufficient level-broadening to allow 
quasiparticles to move along the dispersion relation by hopping from state to 
state.   We refer to Stone (1996) and Kopnin (2002) for details of the 
relationship between the continuum and discrete-$\mu$ cases.  Solution of the 
Boltzmann equation, followed by integration of the rate of change of 
quasiparticle momenta over $\mu$ and ${\bf p}_{\perp}$, gives the force acting on 
the vortex, including the Magnus and spectral flow components.  We do not give 
further details here because the review by Kopnin (2002; pp. 1654 - 1660) 
contains a complete and very transparent account.

On this basis, the force-balance equation per unit length of vortex is,
\begin{eqnarray}
{\bf f}_{B} + {\bf f}_{V} + {\bf F}_{M} + {\bf F}_{l} + {\bf F}_{sf} = 0,
\end{eqnarray}
in which ${\bf f}_{B,V}$ are the external forces caused by buoyancy and 
interaction with neutron vortices.  The Iordanskii force (Thouless, Ao \& Niu 
1996; Sonin 1997) is unimportant here and has been neglected in equation (3).  
The Magnus force is expressed in terms of the proton electromagnetic current 
density ${\bf J}^{s}_{p}$ obtained from the entrainment coefficients (Andreev \& 
Bashkin 1975; Alpar, Langer \& Sauls 1984; Jones 1991), ${\bf v}_{L}$ and the 
superfluid bulk velocities,
\begin{eqnarray}
{\bf F}_{M} = \frac{1}{c}{\bf J}^{s}_{p}\times \phi_{0}.
\end{eqnarray}
A similar expression gives ${\bf F}_{l}$, the electromagnetic interaction with 
electrons and muons, in terms of the lepton current density ${\bf J}_{l}$.  The 
spectral flow force has components parallel with and perpendicular to ${\bf 
v}_{L} - {\bf v}_{T}$,
\begin{eqnarray}
{\bf F}_{sf\perp} = - C\frac{1}{1 + \omega_{0}^{2}\tau^{2}}\frac{\tau}{\tau_{c}}
	\hat{\bf z} \times ({\bf v}_{L} - {\bf v}_{T}),
\end{eqnarray}
\begin{eqnarray}
{\bf F}_{sf\parallel} = - C\frac{\omega_{0}\tau}{1 + 
\omega_{0}^{2}\tau^{2}}\frac{\tau}{\tau_{c}}
	({\bf v}_{L} - {\bf v}_{T}),
\end{eqnarray}
(Volovik 1996; Kopnin 2002), with
\begin{eqnarray}
C = \frac{\pi \hbar \rho_{p}}{m_{p}}\tanh\left(\frac{\beta\Delta_{p}}{2}\right),
\end{eqnarray}
where $\beta^{-1} = k_{B}T$.  The total proton density is $\rho_{p} = 
\rho_{p}^{s} + \rho_{p}^{n}$, but at the temperatures considered here, the normal 
component $\rho_{p}^{n}$ is negligibly small compared with the superfluid 
$\rho_{p}^{s}$.  The simple expressions given by equations (5) and (6) are valid 
at temperatures such that $\omega_{0}^{2}\tau\tau_{c} \gg 1$, which we shall find 
to be always satisfied here.  The definition of relaxation times follows Kopnin 
and Lopatin (1997).

At this point, it is interesting to note that frictional forces on vortices in 
astrophysical superfluids have been estimated previously from the scattering of 
continuum quasiparticles by core quasiparticles, with neglect of spectral flow.
Following this procedure, Sedrakian (1998) found an expression for the frictional 
force on neutron vortices arising from the scattering of continuum proton 
quasiparticles.  In the present context, the order of magnitude of such a force 
would be
\begin{eqnarray}
F^{coll}_{\parallel} = \left(\frac{m^{*}_{p}v_{L}}{\tau_{c}}\right)
\left(\frac{p_{F}k_{B}T}{\pi\hbar\omega_{0}}\right).
\end{eqnarray}
This is smaller than the spectral flow component, equation (6), by a factor of 
the order of $k_{B}T/E_{Fp}$, where $E_{Fp}$ is the proton Fermi energy, and so 
is negligible by comparison. It follows that the frictional force
produced by the scattering of electrons or muons is also negligible because the 
transport relaxation time for this process is several orders of magnitude longer 
than the strong-interaction values of $\tau_{c}$.

\section[]{Application to proton vortices}

Spectral flow is a remarkable quantum phenomenon whose recognition has enabled 
the understanding of laboratory measurements of the force acting on moving 
$^{3}{\rm He}$-B vortices (see, for example, Bevan et al 1997).  Its predictions 
are transparent in the semi-classical modelling of what would otherwise be an 
extremely cumbersome problem.  The ratio of vortex radius to average 
interparticle spacing is given by $p_{F}\xi$ which, for $^{3}{\rm He}$-B, has 
values in the interval $10^{2} - 10^{3}$.  Typical values in the proton 
superfluid are perhaps an order of magnitude smaller.  For example, from the 
energy gap obtained by Baldo \& Schulze (2007; the curve in Fig. 2 of that paper 
containing all corrections) we find $p_{F}\xi = 7.5$ at $\rho_{0} \equiv 0.16$ 
fm$^{-3}$, but this increases rapidly thereafter as a function of matter density. 
Use of the Boltzmann equation does require that the number of quasiparticles be 
large, and this condition is satisfied if we accept that the effective number is 
that contained within a length of vortex at least an order of magnitude greater 
than $\xi$.  The circumstances existing in the core of a neutron star also differ 
from those, for example, of Bevan et al (1997) in relation to the problem of 
defining the heat bath and its velocity.  Apart from collective excitations, the 
heat bath contains the lepton-photon system and continuum quasiparticle 
excitations of the neutron and proton superfluids.  It is also possible that the 
neutrons may exist as a normal Fermi liquid within limited intervals of density.  
All these components interact electromagnetically or strongly with the nuclei 
embedded in the solid crust of the star and, in the absence of vortex motion, 
would be stationary with respect to it.  But any component more strongly coupled 
with moving vortices than with the crust would be expected to flow with them.  
The neutrons are the most important component and in their case we shall follow 
the argument given by Kopnin (2002; Sect. 9 of that paper) in relation to the 
analogous problem in $^{3}{\rm He}$-B and assume that the continuum quasiparticle 
rest frame is stationary with respect to the crust provided their mean free path 
for scattering by other continuum quasiparticles is shorter than the mean free 
path for scattering by localized proton quasiparticles.  In a steady state of 
vortex motion, the force ${\bf f}_{B} + {\bf f}_{V}$ is transferred to the 
neutrons and must be balanced by a neutron chemical potential gradient.  But this 
is very small, of the order of $10^{-6} - 10^{-5}$ eV cm$^{-1}$. 

Equation (4) has been expressed as shown to emphasize that ${\bf F}_{M} + {\bf 
F}_{l} = 0$ because screening of the electromagnetic current in the whole volume 
of the neutron-star core must be presumed for a type II superconductor (Jones 
1991, 2006) except for the microscopic circulating currents of the vortices.  The 
distribution of leptonic current density is present throughout core and crust 
prior to the superconducting transition and in non-superconducting regions, can 
change only on the same long time-scales as the magnetic flux density.  Thus the 
superconducting transition must be to a state of finite supercurrent density such 
as to satisfy the condition ${\bf J}^{s}_{p} + {\bf J}_{l} = 0$ everywhere in the 
core.  In effect, the lepton-proton system is an insulator and the moving-vortex 
induction field induces no large-scale current density.  But ${\bf J}^{s}_{p}$ is 
confined to the core of the star, unlike ${\bf J}_{l}$, and the return currents 
that complete the circuit flow as sheets on the surface of the spherical 
superconducting volume and exclude the magnetic flux present in adjacent 
non-superconducting regions.  In having volume screening, neutron stars differ 
from laboratory superconductors for which screening is primarily a surface 
phenomenon (see Tinkham 1996).

In the low temperature limit, $\omega_{0}\tau \gg 1$, only a small longitudinal 
component of the spectral flow force, given by equation (6), remains.  Owing to 
the cancellation of the Magnus force, equation (3) reduces to ${\bf f}_{B} + {\bf 
f}_{V} + {\bf F}_{sf\parallel} = 0$, showing that movement of proton vortices, 
and hence expulsion of the field, occurs easily under the influence of buoyancy 
forces or of interaction with outward-moving neutron vortices as the rotation of 
the star slows (Jones 2006).  At higher temperatures, the existence of spectral 
flow radically changes this behaviour.

It is convenient to take account of both longitudinal and transverse components 
of the spectral flow force by obtaining an expression for the component of ${\bf 
v}_{L} - {\bf v}_{T}$ parallel with ${\bf f}_{B} + {\bf f}_{V}$.  Provided the 
relaxation times are such that equations (5) and (6) are valid, this component is 
independent of $\tau$ and is,
\begin{eqnarray}
\left({\bf v}_{L} - {\bf v}_{T}\right)_{\parallel} = 
\frac{\omega_{0}\tau_{c}}{C}\left({\bf f}_{B} + {\bf f}_{V}\right).
\end{eqnarray}
Neutrons are the most important heat-bath component.  The relaxation time 
$\tau_{c}$ for interaction with superfluid neutrons, wavevector ${\bf k}$, is not 
well known owing to its exponential dependence on the $^{3}{\rm P}_{2} - 
^{3}F_{2}$ energy gap and our obligatory neglect of the possible existence of 
nodes in this function and of its ${\bf k}$-space anisotropy.  Given these 
limitations, we make the most elementary possible approximation and assume that 
the neutron-proton G-matrix is isotropic and identical with the transition matrix  
for free-particle neutron-proton scattering at an equivalent neutron laboratory 
energy. We obtain
\begin{eqnarray}
\tau_{c} \approx 
\left(\frac{3\beta^{2}\hbar^{3}W}{2m_{p}\bar{\sigma}_{np}}\right)
  \left(\frac{m_{p}}{m^{*}_{p}}\right)\left(\frac{m_{n}}{m^{*}_{n}}\right)^{2},
\end{eqnarray}
in which $\bar{\sigma}_{np}$ is chosen at a laboratory energy equal to the sum of 
the Fermi energies, $E_{Fn} + E_{Fp}$ (for this, the angle-averaged 
centre-of-mass energy in the medium equals that for the free-particle 
scattering).
The function $W$ is $W = 1$ in the case of a normal Fermi system, but for an 
isotropic superfluid is given by,
\begin{eqnarray}
\frac{1}{W} &  =  &  \frac{8}{\pi^{2}}\int^{\infty}_{0}dx\int^{\infty}_{0}dy
\frac{1}{1 + {\rm e}^{-X}}\frac{1}{1 + {\rm e}^{Y}}\frac{1}{1 + {\rm e}^{X-Y}},
\end{eqnarray}
in which $X^{2} = x^{2} + (\beta\Delta_{n})^{2}$, with an identical definition of 
$Y$.
The numerical values of $W$ obtained in the interval appropriate for Table 1 are
satisfactorily fitted by $W = 0.8(\beta\Delta_{n})^{-1}\exp(\beta\Delta_{n})$. 

We adopt effective masses $m^{*}_{p} = 0.8m_{p}$ and $m^{*}_{n} = 0.9m_{n}$ (Zuo, 
Bombaci \& Lombardo 1999) in evaluating equations (1) and (10) and for reference 
purposes, assume a proton fraction of $0.03$ at $\rho_{0}$ (Heiselberg \& 
Hjorth-Jensen 2000) and thus a Fermi energy $E_{Fp} = 7.4$ MeV.  The factor 
relating the dissipative force in equation (9) to the size of the Magnus force is 
$\omega_{0}\tau_{c}$.  The Table 1 evaluations of this are for $p_{\perp} = 
p_{F}$ and for $\Delta_{p} = 0.062$ MeV, selected as approximately the minimum 
value consistent with type II superconductivity at this proton density, which 
gives a reference value $\omega_{0} = 1.23 \times 10^{18}$ rad s$^{-1}$.  At the 
temperatures of the Table, the condition $\omega_{0}^{2}\tau\tau_{c} \gg 1$ is 
satisfied because $\tau_{loc}$ has the same order of magnitude as the 
normal-neutron $\tau_{c}$.  The question of the extent to which the neutron 
quasiparticle rest frame moves with the vortices rather than being stationary 
with respect to the crust can be answered, for the temperatures and $\Delta_{n}$ 
of the Table, by noting that $\bar{\sigma}_{np}$ and the proton-proton cross 
section $\bar{\sigma}_{pp}$ at the equivalent laboratory energy equal to 
$2E_{Fp}$ differ only by a factor of order unity. (See Yao et al 2006; pp. 
339-340. For example, $\bar{\sigma}_{pp} \approx 150$ mb at $20$ MeV, whereas 
$\bar{\sigma}_{np} \approx 100$ mb at $70$ MeV.)  The Kopnin condition is 
approximately satisfied provided the ratio of the neutron continuum quasiparticle 
density to the spatially-averaged density of localized proton quasiparticles 
$(\propto \Delta^{2}_{p}B^{-1}\exp(-\beta\Delta_{n}))$ exceeds unity.  Dragging 
of the heat-bath rest frame is negligible for normal neutrons. In the case of 
superfluid neutrons at the lower temperatures and smaller $\Delta_{n}$, it is 
significant only for spatially-averaged magnetic flux densities $B \sim 10^{14}$ 
G, but is much reduced for $\Delta_{p}$ larger than the minimum.

\begin{table}
  \caption{Values of $\omega_{0}\tau_{c}$ are given in the second column, as a 
function of temperature, for interaction of localized (vortex-core) proton 
quasiparicles with normal neutrons.  The remaining columns give this quantity in 
the case of superfluid neutrons with an assumed isotropic $\Delta_{n}$.  Vacant 
spaces are for temperatures close to or above the critical temperature, or for 
temperatures so low that $\omega_{0}\tau_{c}$ is too large to be of interest.
The assumed value of $\omega_{0}$ is the reference value, $1.23 \times 10^{18}$ 
rad s$^{-1}$.}
  \begin{tabular}{@{}lrrrrr@{}}
  \hline
    $T$     &         & 		&	$\Delta_{n}$  &		   &		\\
   $(10^{8}{\rm K})$ &		&		& (MeV) 		&		&		\\
  \hline
			&  0.00	  &  0.04   &  0.08   &  0.12  &  0.16		\\
  \hline
  0.5  &  420  &      &      &      &      \\
  1.0  &  105  &  1910  &       &       &    \\
  1.5  &   47  &   271  &  2971  &       &    \\
  2.0  &   26  &    94  &   477  &  3240  &   \\
  2.5  &   17  &        &   150  &   641  & 3060  \\
  3.0  &   12  &        &    68  &   212  &  750  \\
  3.5  &  8.6  &        &    37  &    94  &  265   \\
  4.0  &  6.6  &        &    23  &    50  &  119   \\
  \hline
\end{tabular}
\end{table}

The effective component of the vortex drift velocity is given by equation (9) 
with $C = 1.69 \times 10^{10}$ erg s cm$^{-3}$ at matter density $\rho_{0}$, 
neglecting the hyperbolic function.  It is only weakly-dependent on proton-system 
parameters other than $\Delta_{p}$.  Values for the case of stronger proton 
pairing can be found simply by scaling from the Table 1 reference value of 
$\omega_{0}$ using the new values of $\Delta_{p}$ and $E_{Fp}$.  In particular, 
it is to be emphasized that there is no exponential dependence on the value of 
the proton energy gap.

The buoyancy force $f_{B}$ is of the order of $H_{c1}\phi_{0}/8\pi R \sim 1.0$ 
dyne cm$^{-1}$ (Muslimov \& Tsygan 1985; Jones 2006), where $H_{c1}$ is the 
superconductor lower critical field and $R$ is the neutron-star radius.  This 
term can arise because the normal to superconducting transition is so fast that 
the superconductor is in a steady, but not necessarily complete equilibrium 
state.  One aspect of this is that, even for $B < H_{c1}$, the magnetic flux 
cannot be expelled from the proton superconductor (Baym, Pethick \& Pines 1969).  
Even if $f_{B}$ were negligible, the term $f_{V}$ would exist in the case of 
superfluid neutrons, caused by interaction between neutron and proton vortices 
(Muslimov \& Tsygan 1985; Sauls 1989; Jones 1991; Ruderman, Zhu \& Chen 1998).  
It is limited by a critical value $f^{c}_{V}$ at which neutron vortices cut 
through the proton-vortex lattice, whose magnitude is at most of the same order 
as the buoyancy term (Jones 1991).  If the outward radial velocity of neutron 
vortices during spin-down exceeds the velocity given by substituting ${\bf f}_{B} 
+ {\bf f}^{c}_{V}$ in equation (9), they will move by cutting through the proton 
vortex lattice in those cases where a relative sliding motion is not possible.

General predictions are, of course, impossible to make owing to the exponential 
dependence of spectral flow on $\Delta_{n}$.  Together with $\Delta_{p}$, this 
function of density also controls the core neutrino emissivity and hence the 
internal temperature of isolated neutron stars less than $\sim 10^{5}$ yr in age 
(see for example, Fig. 5 of Yakovlev \& Pethick 2004) which lack any substantial 
source of internal dissipation.  It is quite possible that values of $\Delta_{n}$ 
larger than those assumed in Table 1 exist in certain density intervals, but it 
is the regions of small $\Delta_{n}$ that have the greater effect on average 
drift velocities.  The choice of $^{3}{\rm P}_{2} - ^{3}{\rm F}_{2}$ neutron 
energy gaps in the Table is consistent with recent calculations
(Heiselberg \& Hjorth-Jensen 2000; Dean \& Hjorth-Jensen 2003).  At $k_{F} \sim 
2.0$ fm$^{-1}$ and above, higher partial waves make little contribution to 
attractive forces and it is even possible that there are significant intervals in 
which the neutrons are not superfluid.

\section{Conclusions}
It is of some interest that a new quantum phenomenon recognized in relation to 
laboratory studies of vortices in $^{3}{\rm He}$ should have some bearing on 
neutron-star magnetism.
Subject to the reservations described in Sect. 3, the conclusions of this paper 
that follow from the spectral flow phenomenon are as follows.  The presence of 
charged or neutral baryons as a normal Fermi system within some interval of 
core-matter density leads to $\omega_{0}\tau_{c}$ values of the order of those in 
the seond column of Table 1.  These are some orders of magnitude smaller than the 
$\omega_{0}\tau_{c} \approx 5 \times 10^{3}$ needed for velocities $\sim 3 \times 
10^{-7}$ cm s$^{-1}$ that give significant magnetic flux transfer from core to 
crust within the $10^{4}$ yr magnetar phase.  The absence of substantial movement 
of flux from core to crust in this time interval would be
consistent with a field configuration in which the distribution of the active 
flux believed to be responsible for magnetar phenomena is confined to the crust 
at or close to the time of  neutron star formation, its ability to move and 
evolve being unconstrained by linkage with the core.  This is also true, even in 
the absence of normal neutrons, for the $\Delta_{n}$ of Table 1 because the 
higher temperatures there may well be maintained by internal dissipation during 
the active magnetar phase.  Comparison of the continuum and localized 
quasiparticle densities shows that there can be dragging of the heat-bath frame 
${\bf v}_{T}$ by ${\bf v}_{L}$, much reducing spectral flow, particularly for 
very high fields, $B \sim 10^{15}$ G, and for $\Delta_{p}$ near the minimum 
consistent with type II superconductivity, but our broad conclusions are 
unchanged.

There are other circumstances in which spectral flow may be of relevance, such as 
the case in which  the density-dependence of neutron-pairing allows both normal 
and superfluid neutrons within appreciable intervals of matter density. The 
values of $\omega_{0}\tau_{c}$ given in the second column of Table 1 for modest 
internal temperatures $T\approx 10^{8}$ K are small enough to constrain the 
proton vortex drift velocities that are possible in cases of rapid spin-down such 
as the Crab pulsar, and the propeller phase of binary systems.  The outward 
movement of neutron vortices in the superfluid region then occurs by sliding 
relative to the proton vortices or by intersecting them, and is unable to force 
the outward movement of magnetic flux.

\bsp

\label{lastpage}

\end{document}